# DNA toroids form via a flower intermediate

Ryan B. McMillan†, Hilary Bediako†, Luka M. Devenica, Yuxing E. Ma, Donna M. Roscoe, and Ashley R. Carter*

Department of Physics, Amherst College, Amherst, MA, 01002, USA

† Joint Authors.
* To whom correspondence should be addressed. Tel: 1-413-542-2593; Fax: 1-413-542-5235; Email: acarter@amherst.edu

**ABSTRACT**

DNA in sperm cells must undergo an extreme compaction to almost crystalline packing levels. To produce this dense packing, DNA is condensed by protamine, a positively charged protein that loops the DNA into a toroid. Our goal is to determine the pathway and mechanism for toroid formation. We first imaged short-length ($L$=217-1023 nm) DNA molecules in 0-5.0 µM protamine using an atomic force microscope (AFM). At low protamine concentrations (0.2-0.6 µM), molecules dramatically condensed, folding into a flower structure. Dynamic folding measurements of the DNA using a tethered particle motion (TPM) assay revealed a corresponding, initial folding event, which was >3 loops at $L$=398 nm. The initial folding event was made up of smaller (<1 loop) events that had similar dynamics as protamine-induced bending. This suggests that flowers form in an initial step as protamine binds and bends the DNA. It was not until higher protamine concentrations (>2 µM) that DNA in the AFM assay formed small (<10 loop), vertically packed toroids. Taken together, these results lead us to propose a nucleation-growth model of toroid formation that includes a flower intermediate. This pathway is important in both *in vivo* DNA condensation and *in vitro* engineering of DNA nanostructures.

# INTRODUCTION

Sperm cells undergo a dramatic nuclear reorganization during spermatogenesis (1). This nuclear reorganization is needed to create a hydrodynamic sperm head (2, 3), and to protect the DNA from UV damage during travel to the oocyte (4–8). During this reorganization, the histone proteins that wrap the DNA are removed and the genome is compacted by a factor of 40 in some organisms (1, 9).

A key player in the compaction is a family of small (~50-amino-acid), arginine-rich proteins, called protamines (2, 10). Species of vertebrates have 1-15 protamines, with salmon having one protamine (salmine) and humans having two protamines (P1 and P2) (2). Mammals, such as humans, enlist transition proteins to fold the DNA before compaction by protamines, but other animals, such as salmon, use protamine to directly compact DNA (2, 4). For all organisms, DNA compaction by protamine leads to a series of folded DNA toroids (2, 11–13) that hold ~50 kbp of DNA at almost crystalline packing levels (14, 15).

Interestingly, DNA toroids form both *in vivo* and *in vitro*. To form toroids *in vivo*, protamines are initially phosphorylated upon binding the DNA, but afterward, the phosphates are removed and the cysteine residues on nearby protamines form disulphide bonds, stabilizing the toroid (2, 16). *In vitro*, these protein modifications are not required for toroid formation (17, 18). In addition, *in vitro* toroids can be formed by any cation with a charge of at least +3 (19), including metal ions such as cobalt (III) hexaammine (20, 21), naturally occurring polyamines such as spermine and spermidine (22, 23), and other polypeptides such as histone H1 (24) and polylysine (25). These multivalent cations are often called condensing agents (14, 19). Measurements of *in vitro* toroids made by condensing agents show that packing within the toroid is a hexagonal lattice (18, 20), and that toroids have an average inner diameter of approximately 30-50 nm and an average outer diameter of 80-100 nm (12, 18, 26). Since the structure of the toroids formed by different condensing agents is similar (18), it is thought that the mechanism of toroid formation for one condensing agent may be generalizable to all of them.

Here, we are interested in toroid formation by protamine. To form a toroid, positively charged protamine molecules are thought to coat and neutralize (2, 19) the negatively charged DNA (perhaps binding every ~10 bp or so (27) in the DNA groove (9, 28)). Neutralization of the DNA has two effects. The first effect is that neutralization causes the DNA to bend (29). Measurements of bending by protamine hypothesize that each molecule induces a ~20° bend, producing a radius of curvature in the DNA of ~10 nm (30). Multiple bending events (as well as some DNA thermal fluctuations) eventually lead to a loop with a measured 20-35 nm diameter in 105-nm-length DNA (30). The second effect of DNA neutralization is that protamine can cause DNA-DNA interactions that stabilize the DNA, creating loops or toroids. Yet, exactly how protamine uses these two effects (DNA bending and DNA-DNA interactions) to form toroids is unclear.

The current model for toroid formation is a nucleation and growth model (18, 21, 26, 31, 32). In this model, toroid formation is characterized by a nucleation stage in which a single loop forms (perhaps with one DNA-DNA interaction) and a growth stage in which more DNA loops are added to the first one, creating additional DNA-DNA interactions. In this way, the toroid is folded "loop-by-loop", matching single-molecule assays for other condensing agents that show that folding and unfolding happen in discrete, loop-sized steps (33, 34). The problem with this model is that it does not account for DNA bending, which would likely be occurring all along the DNA length and might create several loops at the same time. If the nucleation event for toroid formation is the folding of a single loop, then it is interesting that up to 60 kbp of DNA can form one toroid (13) instead of multiple toroids. Thus, it may be that the nucleation event for toroid formation is not a single loop. Previous studies have found that toroids with 2-6 loops are unstable and spontaneously unfold (34), casting some doubt on a single-loop nucleation event. It is possible that growth of the toroid occurs loop-by-loop, but that nucleation of the toroid follows a different pathway.

Here, we would like to study the nucleation of the toroid by measuring the folding of DNA molecules that form small (<10 loop) toroids. Specifically, we added protamine (0-5.0 µM) to short ($L$ = 217-1023 nm) DNA molecules and either visualized the DNA structures that formed with an atomic force microscope (AFM) or measured the folding in real time with a tethered particle motion (TPM) assay. At low protamine concentrations (0.2 µM), we visualized DNA in our AFM assay that was remarkably condensed (with an extension that was a factor of ~2 less than the nominal value). This condensed DNA was folded into a structure that looked like a "flower" with one or more loops emanating from a common centre. Interestingly, these flower structures have been visualized previously for DNA folded with spermidine (22). We also looked at the real-time folding of the DNA in the TPM assay and noticed multiple steps in the DNA folding that had the same dynamics as protamine-induced bending (30). In addition, there was a large, initial folding event, especially for the longer ($L$ = 398-1023 nm) DNA where we observed folding of >3 loops of DNA. Given this data, we hypothesize that protamine is bending the DNA all along its length, forming a flower intermediate. At higher protamine concentrations (>2 µM), we imaged small (<10 loop) toroids that were vertically packed and a factor of 2 smaller in each dimension than toroids made with longer DNA (12, 18, 20). We thus propose a flower model for toroid formation, in which flowers serve as the intermediate structures between unfolded DNA and small proto-toroids.

## MATERIAL AND METHODS

### Reagents

Protamine from salmon (Sigma Aldrich, P4005) was diluted in deionized water and stored at -20 ºC until ready for use. We also purchased Lambda DNA (New England Biolabs N3011L) and stored at -20 ºC. Anti-digoxigenin from sheep (Roche, No. 11333089001) was diluted to 200 µg/mL in phosphate-buffered saline (pH = 7.4) and stored in 10 µL aliquots at -20 ºC. If this product was unavailable, we instead used anti-digoxigenin from mice (Roche, No. 11333062910).

### DNA construct preparation and protamine

DNA lengths of 217 nm (639 bp), 398 nm (1170 bp), and 1023 nm (3008 bp) were amplified from Lambda DNA using standard PCR procedures (30, 35). We used an LA *Taq* DNA polymerase (TaKaRa Bio RR002) and custom primers (Integrated DNA Technologies). Forward primers were tagged with biotin and reverse primers were tagged with anti-digoxigenin so that the final DNA product could bind to the streptavidin-coated particle and a digoxigenin-coated slide, respectively.

After amplification, we checked the purity of our product using a gel purification procedure. We first performed gel electrophoresis using orange loading dye (New England Biolabs, B7022S) and standard protocols for short DNA molecules (36). Samples were then extracted from the gel using a commercial kit (Qiagen QIAquick Gel Extraction Kit, No. 28704). Finally, we checked the concentration and purity of the PCR product (ThermoFisher, NanoDrop Lite) and discarded samples with an A260/A280 ratio below 1.7.

**AFM sample preparation**

We created AFM samples by affixing mica slides (Ted Pella, grade V1, 10-mm-diameter, ruby muscovite) to magnetic disks using quick-dry epoxy. For control samples, we prepared 20 µL of a solution of 1.0 ng/µL DNA and 2.0 mM magnesium acetate in a DNA LoBind microcentrifuge tube (Eppendorf, No. 022431005) to minimize adhesion to the tube. We pipetted this solution onto the surface of the mica, waited up to 30 seconds, washed with 1 mL deionized water, and dried using nitrogen. Samples were stored in a desiccator.

For all samples with protamine, we found that adding protamine (even small concentrations of 0.2 µM) to 1.0 ng/µL of DNA resulted in DNA aggregates. This is because protamine creates DNA-DNA interactions. To combat this, we reduced the concentration of DNA in solution and performed multiple depositions. Specifically, we prepared 20 µL solutions of 0.2 ng/µL DNA, 2.0 mM magnesium acetate, and the appropriate concentration of protamine. Then, we flowed deionized water to rinse the sample, vacuumed off the liquid, and dried the surface with nitrogen. We repeated this process until there were 2-5 depositions (with 5 depositions working the best). We also did not incubate the solution on the slide, but instead washed the solution off the slide with 1 mL deionized water just after the solution was pipetted onto the mica.

**AFM data collection**

AFM samples were imaged using a Dimension 3000 AFM (Digital Instruments) with a Nanoscope IIIa controller. We used the PPP-XYNCSTR-model cantilever (Nanosensors, resonant frequency = 150 kHz, force constant = 7.4 N/m, length = 150 µm, tip radius < 7 nm). Images were taken in air. There was some DNA shrinkage due to drying in air, but most molecules (70%) were still within 20% of the expected contour length. DNA strands previously imaged with this system show heights of 0.55 ± 0.07 nm and have a lateral radius of 3.9 ± 0.6 nm, close to the expected lateral radius of 3 nm (37). We used a scan rate of either 2-4 Hz. Most images are 2 µm x 2 µm and 512 x 512 pixels, but some are 1 µm x 1 µm and 256 x 256 pixels.

**AFM analysis**

AFM images were processed in Gwyddion (38). We corrected images by (i) aligning rows using a 5[th] degree polynomial, (ii) using the FFT filter and removing high-frequency oscillations, and (iii) removing scars. Using the corrected images, we identified DNA singlet molecules. Singlets had to be lying flat on the surface with at least 1 pixel of separation between other molecules. The contour lengths were required to be within 20% of the nominal length. We cropped square images of valid singlets and saved them as jpegs.

We classified singlets into various categories. An unlooped molecule never crosses over itself and does not bound any area. A looped molecule encloses at least one region of bound area. A flower is a molecule that has one or more loops and contains a central point at which all loops join together. Flowers were subclassified according to the number of loops they possess. Flowers with more than 5 loops were treated as a single category due to the difficulty of precisely counting the number of loops. A molecule with separate loops has multiple loops that are separated and do not contain a central point at which all loops join. A toroid is a structure in which DNA is stacked and circularly symmetric. We also noticed some molecules that did not lay flat on the surface and had regions that were multilooped and regions with significant DNA overlap. These molecules could be collapsing toroids or they could be flowers that landed on the surface incorrectly. We did not study these molecules further.

For all singlets except toroids, we measure three quantities. Extension is the straight-line distance between the two ends of the DNA. If the singlet was too compact to identify both ends, then the extension was the average of two perpendicular measurements across the singlet. Contour length is the distance between the two ends of the DNA measured along the DNA molecule. Loop diameter is measured as the average of two perpendicular diameter measurements. For molecules with multiple loops, we measured the diameter of each loop individually.

To characterize toroids, we computed the diameter $d$ and the maximum height $h$ of the toroid. Specifically, we first measured two perpendicular height profiles using Gwyddion and then imported the profiles into Igor (WaveMetrics, version 6.37). For each profile, we calculated the maximum height of the profile and the profile diameter (e.g. $h_1$ and $d_1$) (Supplementary Figure S1). The profile height is the maximum value minus the minimum value. The profile diameter was found by first identifying all of the local maxima in the profile that were greater than the average height of the profile. If there was only one local maximum in the profile, then the profile diameter was computed as the full width at the location where the height falls off by $1/e$. If there were two or more local maxima, then the profile diameter was set to the lateral separation between the two maxima. Specifically, we computed the outer distance between the locations where the height falls off by $1/e$. The height of the toroid $h$ was set as the average of the two profile heights, $h_1$ and $h_2$. The diameter of the toroid, $d$, was set as the average diameter of the two profiles $d_1$ and $d_2$. Finally, we classified toroids as having a hole or not. A toroid with a hole has at least one profile with 2 local maxima above the average height of the profile. To verify that toroids were comprised of individual DNA molecules, we performed a cut on volume (Supplementary Information, Toroid Volume Cuts and Supplementary Figure S2) because we could not measure the contour length of a toroid.

**TPM sample preparation**

We based our sample chamber preparation procedure off of prior studies (39, 40). To construct the sample chamber, we first adhered a glass coverslip (Fisher Scientific, 12-544-B) to a glass slide (Corning, 2947) using double-sided tape, which creates a ~90-µL microfluidic chamber. In order to stabilize the microfluidic chamber, we applied quick-dry epoxy to the sides and corners of the coverslips.

DNA tether preparation is based on prior studies that used TPM to study DNA folding dynamics (41–43). Specifically, we first prepared a solution of 170 pM streptavidin-coated, polystyrene particles (Spherotech SVP-05-10) that are 560 nm in diameter. We started by washing the particles three times using 0.4% phos-Tween [0.4% Tween-20 in 100 mM Na-Phos solution (pH = 7.5)]. After washing, we resuspended the particles in wash buffer [25 mM Tris-HCl (pH=7.5), 1 mM magnesium acetate, 1 mM sodium chloride, 1 mM dithiothreitol, 0.4% Tween-20, and 3 mg/mL bovine serum albumin] such that the concentration of the particle solution was 340 pM. We then sonicated the particles in a cup sonicator (QSonica, Q500 with Oasis 180 chiller) for 60 minutes using a 2 seconds on, 2 seconds off cycle. After particle preparation, we diluted DNA of the desired length to 200 pM in Tris-HCl buffer (pH=7.5) and combined the particles and the DNA to produce a solution of 170 pM particles and 100 pM DNA. We allowed this solution to incubate for 1 hour to facilitate binding of the streptavidin on the particles to the biotin on the DNA.

Next, we flowed a solution of 20 μg/mL anti-digoxigenin (Roche, No. 11333062910 or No. 11333089001) diluted in 100 mM Na-Phos (pH=7.5) into each sample chamber. After a 1-hour incubation period, we washed each sample twice using wash buffer, waited 10 minutes, and then washed twice again. Finally, we flowed the particle-DNA mixture into the sample chamber and waited for 3-4 hours at room temperature or up to 24 hours at 4 °C. We observed that longer wait times led to more tethers. At the end of the wait period, we washed twice using wash buffer. The result of this procedure is ~10 viable DNA molecules in the field of view that are tethered at one end to the sample surface and at the other end to the particle. Samples can be stored at 4 °C for several days.

**TPM data collection**

We imaged sample chambers using an inverted microscope (Nikon, Eclipse Ti-U) equipped with a 100X, oil-immersion, objective lens (Nikon, CFI Plan Apochromat Lambda Series, numerical aperture = 1.45) and a 12 V, 100 W halogen lamp (Nikon, D-LH/LC Lamphouse). Images were recorded with a CoolSnapEZ camera (Photometrics, full field of view = 1392 x 1040 pixels, camera pixel size = 6.45 mm x 6.45 mm). The image exposure time was 10 ms and the time between images was 200 ms (data rate = 5 Hz). Camera settings were controlled using Micro-Manager 1.4.22, an open source image acquisition software package (44).

To perform a protamine titration experiment, we took videos of a sample chamber under the microscope. For each sample chamber, we started by taking a 2000-frame video of the DNA tethers in wash buffer with no protamine. This is the control trace. Each frame was saved as a 16-bit TIFF file. We then flowed through the buffer with the lowest concentration of protamine, refocused the microscope onto the surface of the sample chamber, and captured 2000 frames. We repeated this step until we reached the maximum protamine concentration.

**TPM data analysis**

Each 2000-frame video was imported into ImageJ (45) and converted into 8-bit binary images using a background threshold. Particles were then tracked using MTrack2 (46), which has a centroid tracking error of 0.28 pixels (47). We set MTrack2 to track all particles that (i) were visible for at least 1800 frames, (ii) moved no more than 50 pixels (2.2 µm) between frames, and (iii) were at least 5 pixels (220 nm) in diameter in order to prevent objects other than particles from being tracked.

Tracks were imported into Igor (Wave Metrics, Version 6.37). First, we converted the particle positions from pixels to nanometres and removed the first-order drift by fitting a line to the position versus time data and then subtracting off that line (48). For control traces, we calculated the standard deviation along each axis over the entire 2000-frame window ($S_x$ and $S_y$) and disregarded traces below an $S_x$ cut off ($L$ = 217 nm had a 60 nm cut off, $L$ = 398 nm had a 90 nm cutoff, $L$ = 1023 nm had a 125 nm cut off) because particles that were stuck to the surface or bound to multiple tethers were likely to have values below this cut off. In addition, we computed the eccentricity

$$e = \sqrt{\left|1 - \frac{S_x}{S_y}\right|}. \qquad \text{(Equation 1)}$$

Theoretically, the particles should be confined to a circular region, corresponding to an eccentricity of 0. Particles bound by multiple DNA molecules will instead be confined to ellipsoidal regions. Thus, we disregard traces for which $e > 0.45$. For traces with protamine, we disregarded all traces that did not have a corresponding control trace or make it through the entire protamine titration series.

For each tether, we analysed the position versus time data along the *x*-direction. We flowed protamine along the *y*-direction. We calculated the time-dependent standard deviation of the position along the *x*-direction, $\sigma_x$, using a rolling 50-frame window. We also calculated a histogram of $\sigma_x$ using a bin of 3 nm, which is about the measurement error for the TPM assay (30). We only display $\sigma_x$ traces for tethers that made it through all protamine concentrations, but we analysed transitions for all tethers.

**RESULTS**

**Protamine forms toroids using a flower intermediate**

In this study, we are interested in the specific pathway and mechanism for how protamine folds DNA into a toroid. Previous studies (21, 33, 34, 49, 50) of condensing agents have favoured a nucleation-growth model of toroid formation that depends on thermal fluctuations in the DNA to create a spontaneous DNA loop. This initial loop is the nucleation event for the toroid. However, we recently found that protamine does not need to wait for thermal fluctuations to cause spontaneous loops and can instead bind and bend the DNA into a loop using ~10 nm radius of curvature (30). Exactly how this bind-and-bend mechanism would affect toroid formation is unclear. One possibility is that loops would form due to protamine bending instead of spontaneous looping and proceed loop-by-loop as before. Another possibility is that protamine would bind and bend everywhere on the DNA all at once, forming multiple loops rather than a single loop. Measuring the initial steps in the toroid formation pathway should give us insight into how protamine coordinates DNA bending during toroid formation.

To look at the pathway for folding DNA toroids, we took AFM images of 398-nm-length DNA molecules at different protamine concentrations (0-5.0 µM) (Figure 1, Supplementary Figures S3 & S4). Many DNA molecules in the images showed aggregation as protamine is able to create DNA-DNA interactions and bridge multiple molecules. We did not study DNA aggregates further. Instead, we used dilute DNA samples (see Materials and Methods) and isolated single DNA molecules for study by creating cropped images (400 nm by 400 nm). Most single molecules [93% (91 out of 98 molecules)] were unfolded in the absence of protamine and had no spontaneous looping. However, when 0.2 µM protamine was added, we saw that the DNA started to bend and curve, with 20% of molecules remaining unlooped, 24% forming single loops, 42% forming multiple loops, and 4% forming toroids. The presence of these structures with multiple loops is particularly intriguing since this observation would not be predicted by toroid formation models with a single loop nucleation event. Instead, these multiloop structures are predicted in the bind-and-bend model for protamine-induced DNA folding. As we increased the protamine concentration to 5.0 µM protamine, the number of toroids increased to 17%, indicating that we indeed are measuring the toroid formation pathway.

When we examined the multiloop DNA structures in greater detail, we noticed that 13% of molecules had multiple loops that were separate (Supplementary Figure S5), while the majority (87%) had DNA structures that looked like flowers. Specifically, these flowers had loops or "petals" that shared a common DNA overlap point or "receptacle". Flowers have been observed previously in AFM images of DNA condensed with spermidine (22). Here we expand the definition of flowers to include DNA folded into 1, 2, 3, or even more loops to emphasize that all these flower structures have a single DNA overlap point. We hypothesize that flowers are formed from protamine bending all along the molecule (which could create one or more loops), as well as DNA-DNA interactions that create the DNA overlap. We note that flowers in solution may look different than flowers immobilized on the surface. It is possible that flowers in solution appear as a long DNA coil with some DNA-DNA interactions.

To quantify DNA folding and toroid formation, we examined the extensions of single DNA molecules across three different lengths (217 nm, 398 nm, and 1023 nm) and multiple protamine concentrations (0-5.0 µM) (Figure 2A). The extension of the DNA is the distance between the two DNA ends, which should decrease as the molecule becomes more folded. Without protamine, the unfolded molecules had average extensions of 144 ± 5 nm, 209 ± 9 nm, and 240 ± 20 nm for $L$ = 217 nm, 398 nm, and 1023 nm, respectively. When we add 0.2 µM protamine, we observed a large decrease (factor of ~2) in the average extension to 88 ± 4 nm, 88 ± 6 nm, and 100 ± 10 nm for the three lengths, respectively. This indicates that there is a large DNA folding event upon the addition of small amounts of protamine, consistent with protamine bending the DNA all at once.

Another quantification of DNA folding and toroid formation is to count the number of molecules of each structural type (*e.g.* unlooped, flower with 1-5 loops, or toroid) at the different protamine concentrations. To do this, we classify every DNA molecule (Figure 2B) and then turn this count into a fractional value by dividing the count by the total number of molecules at each DNA length and protamine concentration. For 217-nm-length DNA, we see that the number of 1-loop and 2-loop flowers at 0.2 µM protamine is ~10% and then grows to ~40% and ~20% by 2.0 µM protamine, respectively. Toroids aren't present until 3.5 µM protamine. For 398-nm-length DNA, we see that ~65% of molecules are flowers with 1-3 loops at protamine concentrations of 0.2-2 µM protamine. Toroids aren't present in large numbers until 5.0 µM protamine when they are ~30% of molecules. Similar results hold for 1023-nm-length DNA, where 64% of molecules are in 2-4 loop flower structures. Toroids are not present until 3.5 µM protamine. This data suggests that even relatively low protamine concentrations (0.2-2.0 µM) create multiloop flowers, rather than single loops or toroids. Increasing the protamine concentration, increased the average number of loops observed in singlets (e.g. 1.5 ± 0.1 at 0.2 µM versus 2.0 ± 0.2 at 2.0 µM for $L$ = 398 nm).

Given this data, we suggest that toroid formation does not involve a single-loop nucleation event, but instead involves a flower intermediate with one or more loops. The formation of this flower is consistent with a bind-and-bend model for DNA, as protamine bends the DNA all at once into the flower. The other main structures in the pathway for toroid formation are an unlooped DNA molecule and a small (<10 loop) toroid. To better characterize this pathway, we will need to study the folding dynamics for toroid formation and the pathway for how flower intermediates form.

**Protamine forms toroids using an initial, large folding event**

Having characterized the structures involved in toroid formation, we then wanted to measure the folding dynamics for the toroid. We anticipated three possible outcomes. One outcome, given the loop-by-loop growth of toroids previously observed (33, 34), is that we see folding happen one loop at a time. A second outcome is that we see multiple steps in the folding that correspond to DNA bending, which might happen at lengths of less than a loop. Finally, a third outcome is that protamine folds the DNA in a single step. Unfolding studies have found that small (2-6 loop) toroids folded by spermine are unstable and unravel all-at-once (34). If folding proceeds in a similar manner to unfolding, we might expect single-step folding. Real-time folding measurements would be needed to differentiate between these possible outcomes.

To measure the folding dynamics for toroids, we use the tethered particle motion (TPM) assay (Figure 3A). In TPM, we track the motion of a 0.56-µm-diameter particle tethered to the surface by a DNA molecule. When we add protamine, the DNA progressively folds into a toroid, constraining the motion of the tethered particle and giving a real-time readout of the folding. This assay is particularly useful for measuring folding by protamine because it is performed in the absence of an external force. Here, we prepared TPM slides using 217-nm-length DNA and titrated in protamine at concentrations ranging from 0.1–0.4 µM (Figure 3B). We tracked the position of one tethered particle over 2000 frames at 5 Hz for each protamine concentration (Figure 3B). We observed that the range of motion of the tethered particle became more constrained at higher protamine concentrations. To quantify this decrease, we computed the standard deviation of the particle over time using a rolling 50-frame window (Figure 3C) since standard deviation is a readout of DNA contour length (Supplementary Figure S6) and is a similar measurement to extension in the AFM assay (Supplementary Figure S7). A histogram of the standard deviation for each concentration reveals a peak at each of the folding states for the molecule (Figure 3D). Specifically, we identified peaks as local maxima with at least 50 counts. We repeated this peak-finding algorithm for all 217-nm-length DNA tethers (N=34) and display the results in a histogram (Figure 3E). We used the same analysis procedure to analyze 398-nm-length (N=21) and 1023-nm-length (N=18) DNA, plotting the standard deviation traces from individual DNA tethers and the locations of the peaks from all of the DNA tethers (Figure 4).

In all three DNA lengths, we observe very similar dynamics. As we increase the protamine concentration, we observe the dynamic folding of the DNA molecules to smaller DNA lengths. During this folding, we see multiple states that are long-lived, on the order of 10s-100s of seconds. The transitions between states are discrete, lasting ~1 s or less (Supplementary Figure S8). We also see both forward transitions indicating folding and reverse transitions indicating unfolding (Figure 4A, Supplementary Figure S8-S9). These folding dynamics involving multiple, discrete, reversible steps are very similar to dynamics we observed for protamine-induced bending on 105-nm-length DNA that only forms a single DNA loop (30).

However, there is an important difference in the folding dynamics we observe here as compared to the dynamics for loop formation. Namely, that the amount of folding was larger for the longer molecule. For example, 1023-nm-length DNA had 8-110 nm steps in $\sigma_x$ (9-945 nm changes in length, see Supplementary Figures S6 & S8). Molecules did not show only step sizes of <1 loop as seen in 105-nm-length DNA molecules (30). This means that toroid folding includes step sizes that are smaller than a loop and step sizes that are larger than a loop. It may be that binding or bending of the DNA is cooperative to produce the variation we see. An alternate explanation is that protamine is stabilizing spontaneous fluctuations in the longer DNA. Still, another explanation is that the spatiotemporal resolution of our instrument [<10 nm in $\sigma_x$ over a bandwidth of 0.005-0.05 Hz and for $L$ < 500 nm (30)] is not able to visualize the bending from individual protamine binding events. This is the most likely explanation, as performing the assay with smaller protamine concentrations (0.03 µM) produced more steps in the data, especially at the longer lengths.

Interestingly, we do not see peaks in the standard deviation traces that correspond to folding intervals of a single loop. Instead, we see a single large initial folding event. To see this clearly, we overlay dashed lines at standard deviations that correspond to 1, 2, 3, or more loops (Figure 4B). Calculation of the locations of these dashed lines is done by assuming a particular DNA length for a loop (Supplementary Information, Calculating Average Loop Diameter) and using a calibration curve to convert between DNA length and standard deviation in the TPM assay (Supplementary Figure S6). Rather than a regular pattern of folded states at intervals of a loop, we see a peak in the standard deviation trace corresponding to an unfolded molecule and then another peak (or several peaks) at states where the molecule is almost fully folded ($\sigma$ < 50 nm). This change in the standard deviation means that even at a low protamine concentration of 0.2 µM, 85% of 217-nm-length DNA molecules fold past 2 loops, 95% of 398-nm-length DNA molecules fold past 3 loops, and 100% of 1023-nm-length DNA molecules fold past 7 loops.

Thus, we did not observe any of the expected outcomes for folding dynamics for the toroid. We see no evidence of regular folding at loop-sized intervals in either the individual traces or the ensemble measurement. We therefore find it unlikely that the toroid is formed by stacking loops of DNA on top of one another loop-by-loop. Instead, we see an initial, multiloop folding event that is likely the formation of the flower. This initial folding event is made up of smaller steps (at lower protamine concentrations like 0.03 µM) that have similar dynamics as the dynamics for folding a single DNA loop. We suspect that flower formation is similar to loop formation in that protamine is binding and bending everywhere all at once, creating an initial dramatic folding event. We also observe that the DNA can be folded to fractions of a loop, creating molecules that are completely folded all the way down to the resolution limit of the instrument ($\sigma_x$ = 6 nm).

**Flowers form through both spontaneous looping and protamine-induced bending**

To study the physical mechanism for the formation of the flower intermediate in more detail, we wanted to determine whether looping in the flower was due to spontaneous fluctuations in the DNA that would be stabilized by DNA-DNA interactions (18, 19, 30) or it was due to protamine-induced bending (30). To differentiate between these two possibilities, we observed the sizes of the loops within the flower intermediate (Figure 5). Protamine-induced bending has a radius of curvature of ~10 nm, forming 20–35-nm-diameter loops (30), while spontaneous folding of the DNA creates loops of many different diameters (17-176 nm in 1023-nm-length DNA). Therefore, measuring loop diameter gives insight into the mechanism of loop creation.

We measured the diameter of all DNA loops in molecules identified as flowers (*N* = 124 for 217-nm-length DNA, *N* = 363 for 398-nm-length DNA, *N* = 193 for 1023-nm-length DNA) (Figure 5A). We observed peaks in diameter at 26 ± 3 nm in 217-nm-length DNA, 25 ± 5 nm in 398-nm-length DNA, and at 32 ± 8 nm in 1023-nm-length DNA. These peak diameters are consistent with protamine-induced looping of 20-35 nm (30). Distributions were right skewed, with the majority of loops <35 nm in diameter (91% in 217-nm-length DNA, 75% in 398-nm-length DNA, and 59% in 1023-nm-length DNA). We would expect this tight range on loop diameter over many DNA lengths if protamine-induced bending of the DNA is playing a major role in the formation of the flower intermediate. However, we also noticed that the average loop diameter decreased with the number of loops. Thus, we wondered if the finite length of the DNA might be playing a role as well.

To tease apart the role of DNA geometry from the role of spontaneous looping and protamine-induced bending, we plotted the diameter of the loop as a function of the number of loops (Figure 5B). In this plot, we again see that the majority of loops, regardless of DNA size or number of loops in the molecule, have diameters <35 nm. However, we also see a decrease in the upper bound on the loop diameter as the number of loops in the molecule increase.

To look at this more closely, we overlay three models onto our data that set geometric limits on loop diameter (Supplementary Information, Geometric Limits on Loop Diameter). In the first model, we consider the absolute maximum loop diameter for an *n*-loop flower. This should occur when *n*-1 loops have diameters at the detection limit of the instrument (3.9 nm) and the remaining loop contains the rest of the DNA length. When we examined this upper bound, we saw that no experimentally observed loop diameter approached it. We noticed that the experimental data on loop diameter appears to follow a 1/*n* scaling law. A simple model that has this property is one in which all loops are exactly the same size and the entire length of the DNA is in a loop. This model mostly predicted the upper bound on loop diameter in experimental data, but some loops (2% in 217-nm-length DNA, 3% in 398-nm-length DNA, and 1% in 1023-nm-length DNA) were outside the model's predicted upper bound. We might set a more accurate upper bound on loop diameter by relaxing the assumption of uniform loops. Specifically, we allow one loop to be larger than all of the others by a heterogeneity factor $H$. Here, $H$ is set to 1.75 since the heterogeneity (largest loop diameter divided by smallest loop diameter) for measurements of single loops is this value (30). This model predicts a slightly higher upper bound on loop diameter that fewer loops exceeded (1% in 217-nm-length, 1% in 398-nm-length DNA, and none in 1023-nm-length DNA).

Now if we look at our data in relation to the geometric limit set by non-uniform loops on a finite DNA strand, we can more clearly resolve looping due to protamine-induced bending versus looping due to spontaneous DNA fluctuations. Within this geometrical limit, protamine-induced bending is still setting the diameter for the majority of the DNA loops measured. However, the range of possible diameters is the full possible range, indicating that spontaneous looping must be playing a role. Perhaps each protamine molecule induces a bend with a particular radius of curvature, but the number of protamine molecules required to bind to the DNA to form a loop varies due to spontaneous looping. This process would create the array of loop sizes we observe.

**Toroids initially form via vertical packing**

The DNA within large, ~100-nm-diameter toroids is arranged in a hexagonal lattice (with some defects) according to cryo-electron microscopy experiments that image cross sections of DNA toroids formed by cobalt (III) hexaammine (20). We wondered if small (<10 loop) toroids are also hexagonally packed or if some other arrangement is more favourable. Hexagonal lattices have been used to model packing of toroids of 6 loops or more (34) since hexagonal packing maximizes DNA-DNA interactions (18). However, it is possible that the energetic benefit of these DNA-DNA interactions may be less dramatic when there are fewer DNA loops. For small (<10 loop) toroids, there may be some benefit to minimizing defects in the toroid rather than maximizing DNA-DNA interactions. Defects in the toroid occur because the DNA is one long molecule, eliminating the ability of the DNA to lie completely flat in the toroidal crystal. If there is an energetic gain to minimizing defects for small toroids, then growth might occur exclusively in the horizontal direction (horizontal packing) or in the vertical direction (vertical packing).

To determine which packing is most likely, we first measured the heights and diameters of toroids made from short ($L$ = 217-1023 nm) DNA (Figure 6). For all of the toroids imaged, we computed the diameter $d$ and height $h$ (Figure 6A). Toroids had diameters of 20–40 nm and heights of 2–5 nm, which were a factor of 2–4 smaller than toroids made from longer DNA (12, 18, 20). As the DNA length decreased, $d$ and $h$ decreased, and the number of DNA loops in the toroid decreased from ~8 loops at the highest DNA length to 2-3 loops at the shortest length of 217 nm. For many toroids, we cannot visualize the hole in the middle of the toroid. Since this occurred for toroids with smaller diameters and larger heights (Figure 6A), we hypothesize that this may be due to the lateral resolution of our instrument since the finite radius (7 nm) of the AFM tip is close to the measured radius of curvature (~10 nm) for protamine-induced bending.

After measuring the height and diameter of the toroids, we compared these measured values to the theoretical predictions of three packing models: horizontal packing, vertical packing, and hexagonal packing (Figure 6B, Supplementary Table S1, and Supplementary Information, Toroid Packing Models). In horizontal packing, the toroid height should be constant as the diameter increases due to loops adding outwardly. In vertical packing, the DNA stacks one loop on top of the other, so that if a DNA molecule with a finite length forms a toroid with a decreased diameter, then it should have an increased height. This leads to height scaling linearly with the inverse diameter. Finally, in hexagonal packing, there is a discrete set of possible diameter-height pairs given by the stacking in the toroid. When we compared the measured toroid properties to these three models (Figure 6B), we observed that, for 398-nm-length DNA, 63% of the toroids agreed with the vertical packing model within error and an additional 30% did not agree with any model, but were closest to the vertical packing model. To verify that this observation holds across the entire range of lengths tested, we rewrote the vertical packing model such that the DNA length $L$ was not a parameter for the model and then plotted the pooled datasets from all three lengths against the vertical packing model (Figure 6C). Data displayed a positive, linear relationship (with correlation coefficient $r = 0.83$), as predicted.

Our data suggests that small (<10 loop) toroids exhibit vertical packing. Vertical packing minimizes defects in the toroid lattice and may be more energetically favourable at small numbers of loops. However, vertical packing suffers from a small number of DNA-DNA interactions (18), which may be the reason why 2–6 loop toroids have been found to be unstable in unfolding experiments (34).

**DISCUSSION**

Our goal was to determine the pathway and dynamics for toroid formation. We hypothesize that toroid formation follows a nucleation-growth pathway with a flower intermediate (Figure 7).

Specifically, we hypothesize that the first step in this toroid formation pathway is the bending of the DNA into a flower structure. TPM measurements that measure folding dynamics show an initial folding event that is larger than a single loop of DNA at low protamine concentrations (0.2 µM). Likewise, AFM measurements of static structures at the same protamine concentrations show an initial decrease in the extension of the DNA that corresponds to DNA molecules folded into a flower intermediate. This flower intermediate appears on the surface as a flower with one or more loops emanating from a central region with DNA-DNA interactions, but the flower in solution might just be DNA bent into a structure with a ~10 nm radius of curvature. Loops within the flower have an average diameter of 25-32 nm, consistent with protamine-induced bending (30). However, the range of measured loop diameters spans the possible values given the finite length of the DNA, indicating that flowers are also, in part, formed by spontaneous DNA looping.

This flower intermediate is probably not the nucleation event for the toroid, because it is unstable. Measurements of folding dynamics on short (217–1023 nm) DNA molecules with TPM show both forward and reverse events, indicating folding and unfolding of the molecule. These unfolding events could be the breaking of DNA-DNA interactions or the unbending of the DNA. We speculate that the nucleation event is likely to be the collapse of the flower to a proto-toroid.

One candidate for the proto-toroid is the small (<10 loop) toroid we observe here. Measurements of the heights and diameters of these toroids show that their dimensions are about a factor of 2 smaller than the larger toroids with 50 kbp of DNA. These measurements of height and diameter also indicate a vertical packing within the toroid, rather than the hexagonal packing observed for toroids on longer DNA molecules (20). This difference in packing suggests different formation pathways with small (<10 loop) toroids forming by collapse of the flower intermediate and larger toroids forming during the growth stage. During toroid growth, DNA loops would add to the toroid, perhaps one at a time (34), using hexagonal packing.

Here, we did not explore the pathway for how the flower intermediate collapses into the proto-toroid. The TPM assay does not distinguish between DNA folded into a flower versus the same amount of DNA folded into a toroid. It only measures the change in the DNA length of the tether. In the AFM assay, it is difficult to image structures during the collapse of the flower. Future force spectroscopy assays could look at the collapse of the flower intermediate into a toroid.

We note that a flower intermediate in toroid formation may be well-suited for protamine's role in sperm cells (51). For example, if the nucleation event is the collapse of the flower intermediate to a proto-toroid, then aberrant single loops would not trigger toroid formation. In addition, the introduction of a flower intermediate adds an additional level of robustness to the DNA condensation process, giving the cell more opportunity for regulation, perhaps with protein modifications of protamine (16). Finally, after fertilization when protamines are removed (51), the instability we measure here for flowers (and possibly small toroids) might be advantageous.

Given the similarities of DNA toroids formed from different condensing agents (18, 22), it is likely that the pathway we describe here for protamine is more general. We speculate that this nucleation-growth pathway with a flower intermediate may apply to toroid formation by all multivalent cations. If this is true, then these multivalent cations might be useful in triggering the rapid condensation of DNA assemblies, for example in a DNA origami nanostructure (52).

## SUPPLEMENTARY DATA

Supplementary Data are available at NAR online.

## FUNDING

This work was supported by a Cottrell Science Award from the Research Corporation for Science Advancement (ARC, Award #23239), a CAREER award from the National Science Foundation (ARC, Project #1653501), and Amherst College. Funding for open access charge: National Science Foundation Project #1653501.

**FIGURES**

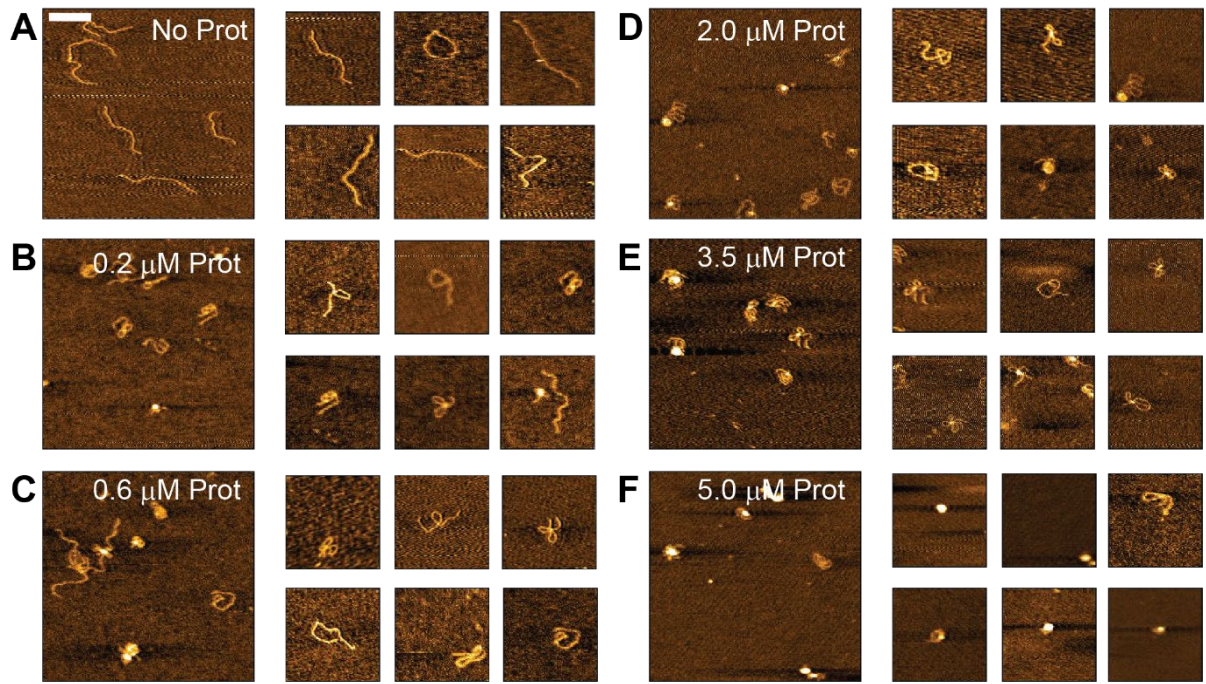

**Figure 1: AFM images show flower intermediates.** A-F) Representative 1 μm x 1 μm AFM height scans of 398-nm-length DNA (*left*) and representative singlets (*right*) at each protamine concentration are shown. Scale bar is 200 nm.

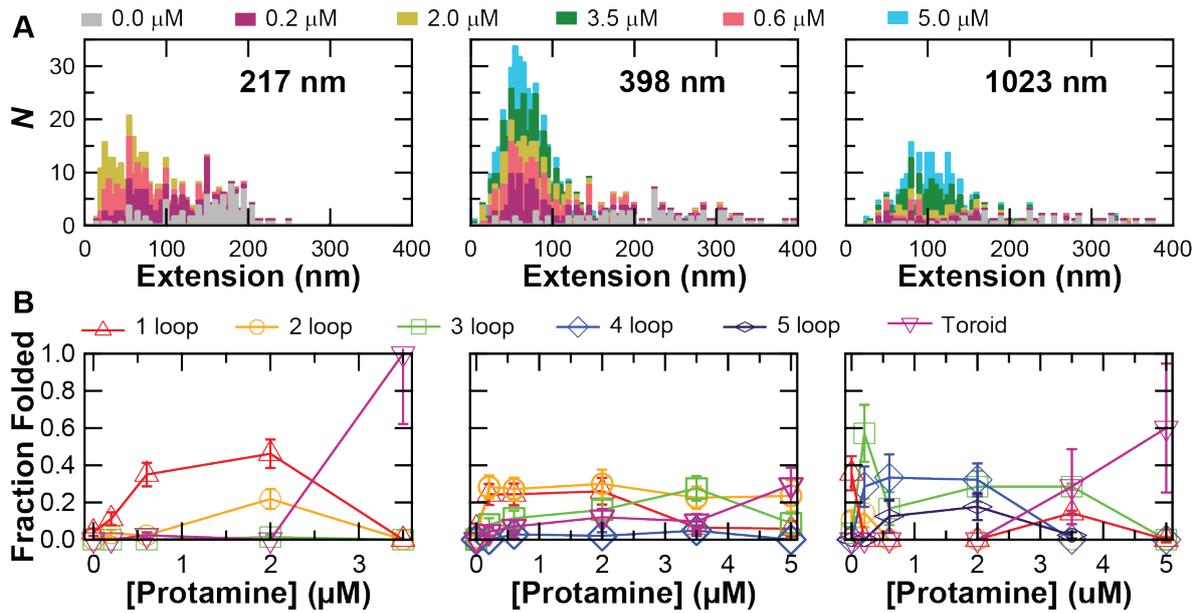

**Figure 2: AFM measurements vary with protamine concentration.** A) Histograms of the extensions of AFM singlets and (B) fraction of all singlets in a particular conformation at each protamine concentration for 217-nm-length DNA (*left*), 398-nm-length DNA (*centre*), and 1023-nm-length DNA (*right*). Histograms are stacked and bin size is 5 nm. Error bars are calculated using Poisson statistics. Only flower structures are counted towards the fractions, not separate loop structures. Colour scheme in legend.

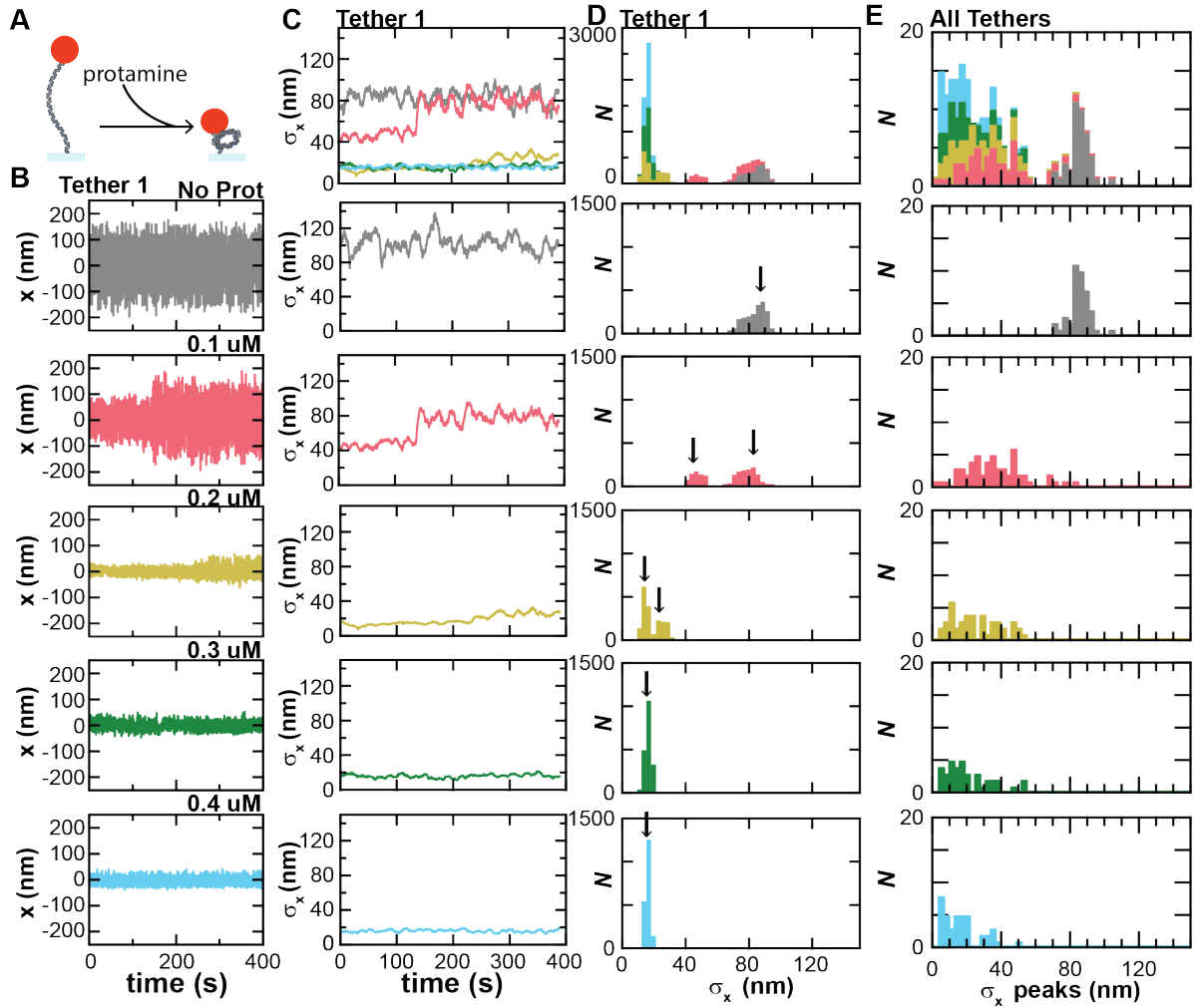

**Figure 3: TPM dynamics show multiple, long-lived states.** A) Cartoon depiction of the TPM assay. Protamine causes the DNA tether to fold, which restricts the motion of the attached particle. B) Position versus time data over a series of 400-second traces at various protamine concentrations for a 217-nm-length DNA tether. C) We compute the standard deviation of the position σ$_x$ over a rolling 50-point window and summarize the data both as a compilation (*top*) and as individual traces (*bottom*). D) Histograms of the standard deviation data reveals that this tether occupies at least 5 distinct states (*arrows*). E) We repeat the process shown in parts B-D for all tethers, calculate the local maxima for each histogram, and then display the data in a histogram. All histograms are stacked. Bin size is 3 nm.

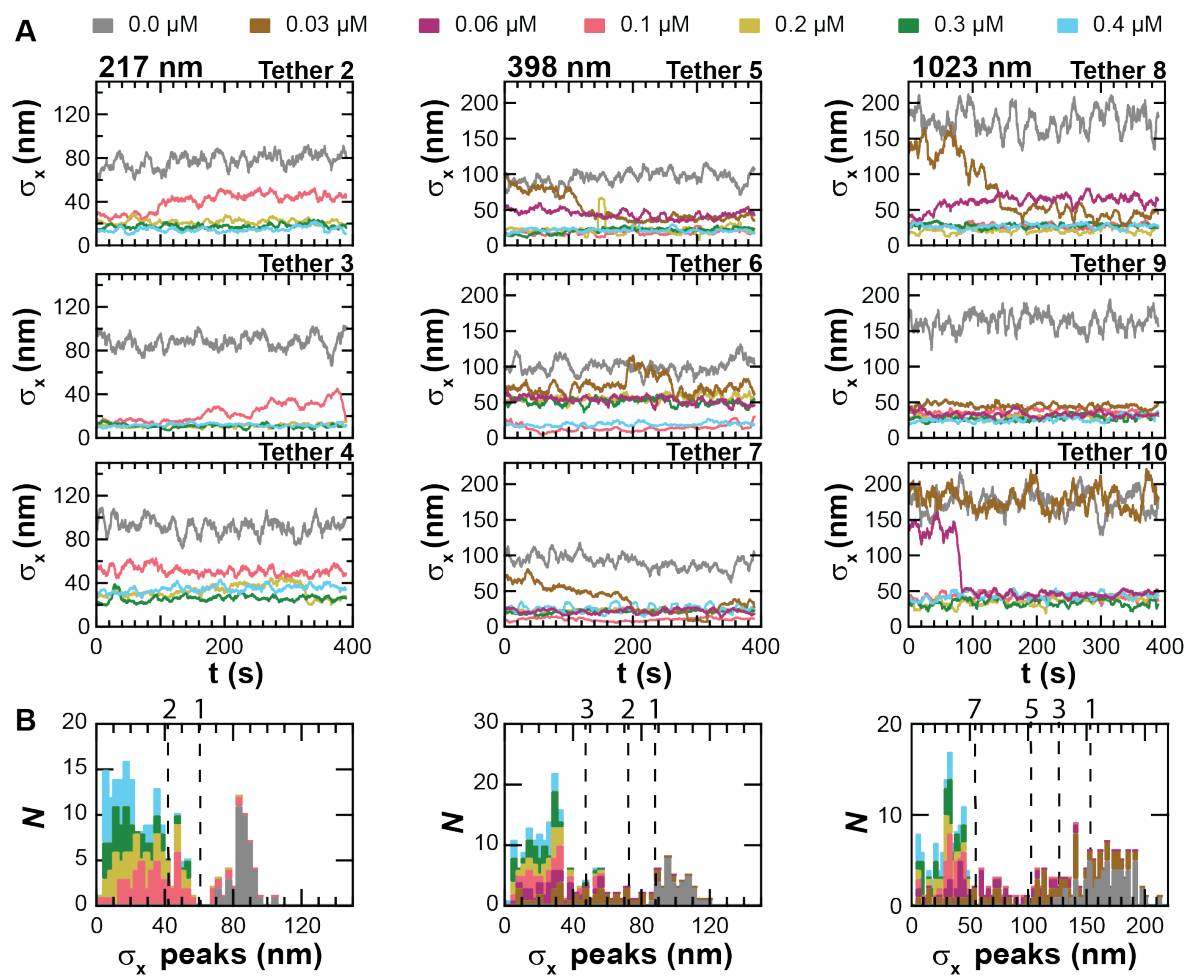

**Figure 4: TPM dynamics at three different DNA lengths show a large initial folding event.** A) Plots of the rolling standard deviation $\sigma_x$ at each concentration over time for three individual 217-nm-length DNA tethers (*left*), three individual 398-nm-length DNA tethers (*centre*), and 3 individual 1023-nm-length DNA tethers (*right*). A decrease in $\sigma_x$ indicates DNA folding. Data is at 0.1 Hz. B) Histograms of $\sigma_x$ for all of the 217-nm-length, 398-nm-length, and 1023-nm-length DNA tethers. Computed locations of different looped states are marked (*dashed lines*). Histograms are stacked one on top of the other. Color scheme is the same as that in Figure 3. Bin size is 3 nm.

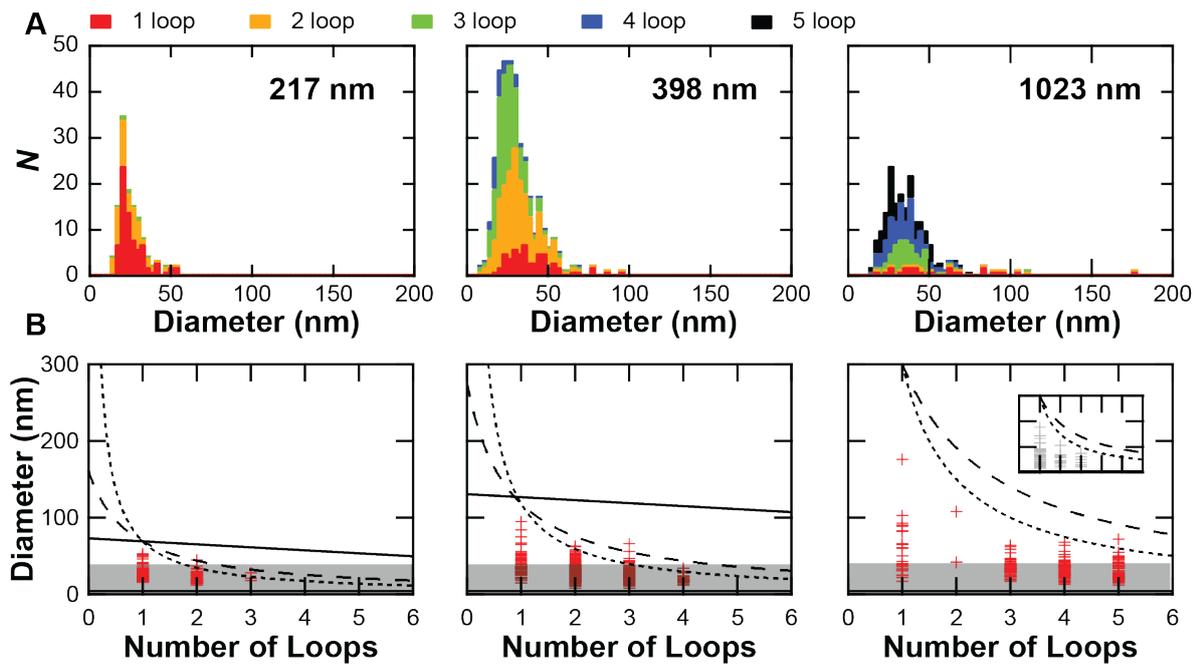

**Figure 5: Loops in flowers form via a combination of protamine-induced bending and spontaneous thermal fluctuations.** A) Histograms of the diameters of individual loops in flowers for 217-nm-length (*left*), 398-nm-length (*centre*), and 1023-nm-length (*right*) DNA. Bin size is 3 nm. Histograms are stacked. B) Diameters of all loops observed in 217-nm-length DNA (*left*), 398-nm-length DNA (*centre*), and 1023-nm-length DNA (*right*) are plotted against the number of loops in the flower. Inset shows 1023-nm-length spontaneous loops on the same axes as the other plots. Distribution is similar to that observed in experimental data. Maximum and minimum bounds are shown (*solid lines*), along with fits from a model that assumes uniform loops (*small dashes*) and another that assumes heterogeneous loops (*large dashes*). The region of protamine-induced loops is shaded.

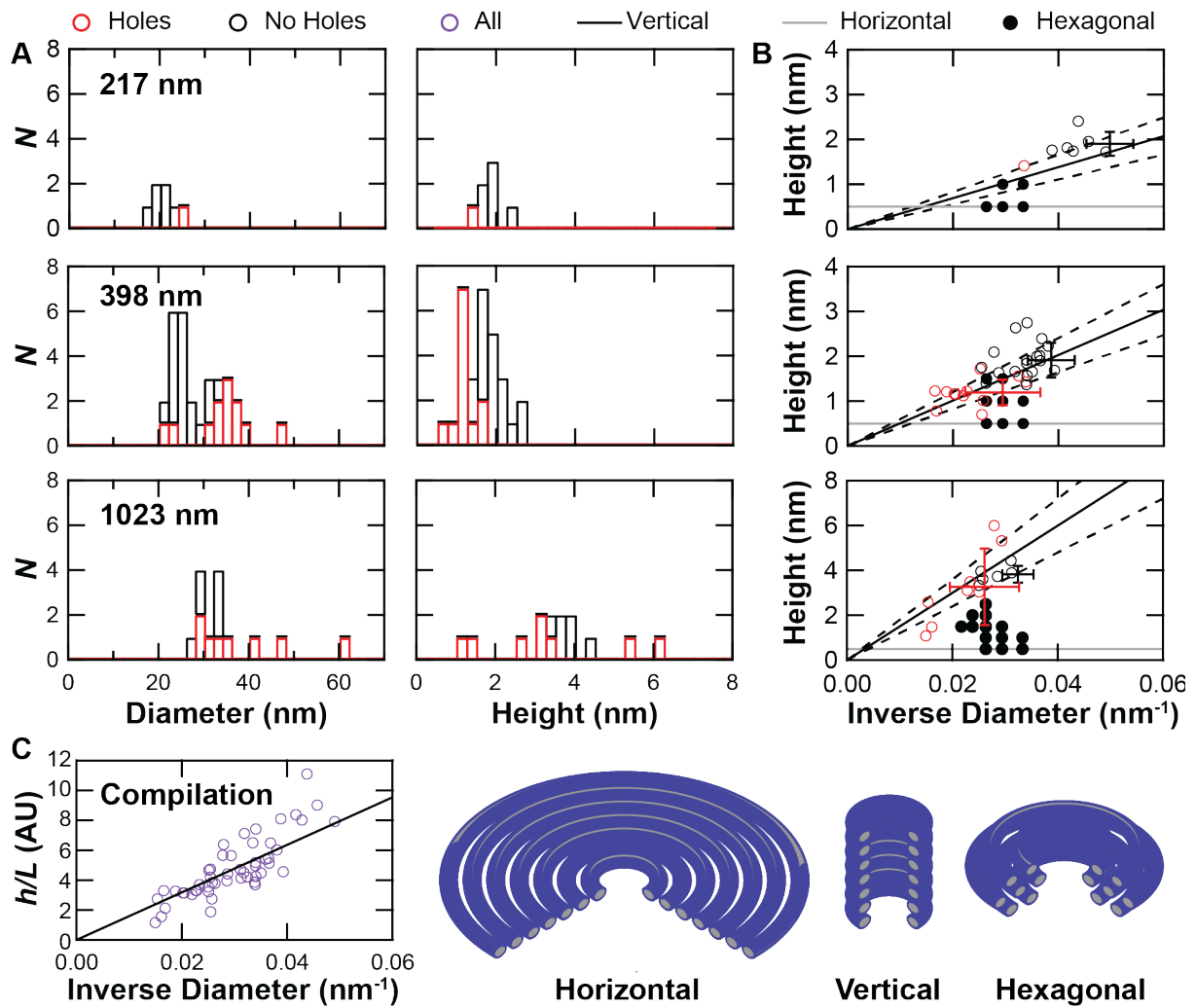

**Figure 6: Protamine folds DNA into vertically packed toroids.** A) Histogram of toroid diameters (*left*) and heights (*right*) for toroids formed from 217-nm-length DNA (*top*), 398-nm-length DNA (*centre*), and 1023-nm-length DNA (*bottom*). Histograms for toroids with holes (*red*) and toroids without holes (*black*) are stacked. Bin size is 2 nm for diameter and 0.25 nm for height. B) Plots of height vs. inverse diameter for toroids from 217-nm (*left*), 398-nm (*centre*), and 1023-nm (*right*) DNA. Toroids are separated into those with holes (*red hollow circles*) and those without holes (*black hollow circles*). Average dimensions with standard deviations are shown for both toroids with holes (*red cross*) and toroids without holes (*black cross*). Predictions from the horizontal packing model (*grey solid line*) and the vertical packing model (*black line*) are shown. The vertical packing model includes 20% error on the height of the DNA (*black dashed lines*) because the model is particularly sensitive to this parameter. The hexagonal packing model comprises a set of discrete points (*black points*). Below are cartoons of the three toroid packing models. C) We plot the height divided by the contour length for all toroids (*purple hollow circles*) against the inverse diameter and compare to the vertical packing model (*black line*).

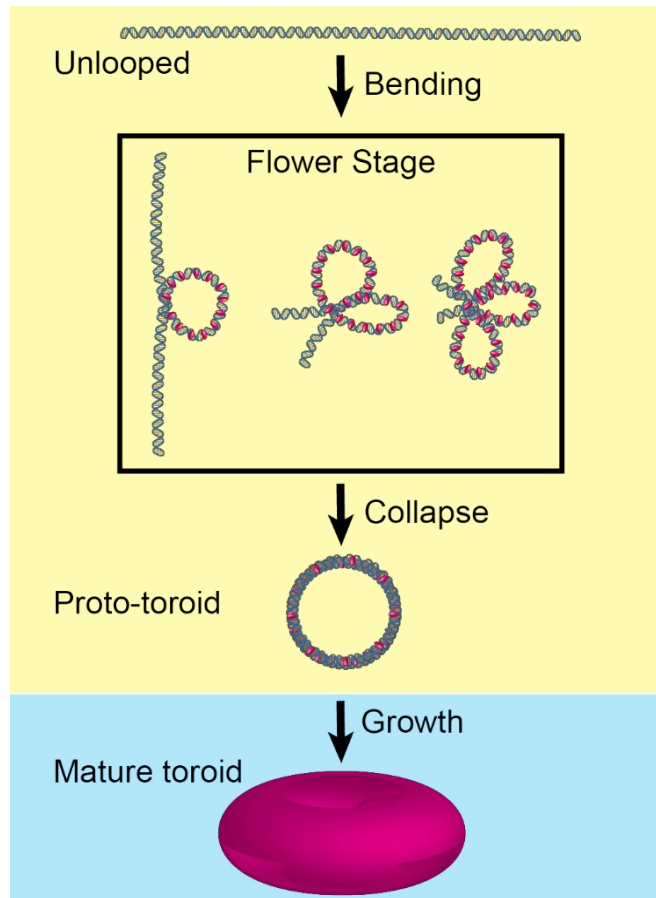

**Figure 7: Toroid formation pathway with a flower intermediate.** During the nucleation stage (*yellow*), protamine binding induces bending along the DNA molecule, creating flower structures with various numbers of loops. The DNA then condenses into a proto-toroid. This proto-toroid transitions into the growth phase (*blue*) as more DNA loops are added, eventually forming a mature toroid.

## Supplemental Information

### Toroid Volume Cuts

To determine which toroids were formed from individual DNA molecules, we first calculated the predicted volume of a toroid comprised of DNA of the relevant contour length, *L*. We assumed that toroids are made of 30-nm-diameter loops stacked one on top of the other. The height of each loop should be 0.5 nm (a typical value for the height of DNA imaged using AFM) (1). The total height of the toroid is then 0.5 nm times the number of loops. The number of loops is $\frac{L}{\pi(30\ nm)}$. We modeled the toroid as an ideal cylinder, for which the expected volume is:

$$V_{expected} = \pi(15\ nm)^2 \frac{L}{\pi(30\ nm)}(0.5\ nm) \qquad \text{(Equation S2)}$$

Next, we computed the calculated volume of a toroid by revolving its cross-sectional area about its central axis. We estimated the cross-sectional area by numerically integrating the two height profiles to yield two cross sectional areas, $A_1$ and $A_2$, and then averaging the two cross sectional areas together. Thus, for a toroid with measured diameter *d*, the calculated volume is:

$$V_{calculated} = \frac{A_1 + A_2}{2} \pi d \qquad \text{(Equation S3)}$$

We analysed toroids for which $0.5 V_{expected} < V_{calculated} < 1.7 V_{expected}$. To set the lower bound, we reasoned that toroids might have shrunk and become compressed vertically such that their volumes could be as small as half the expected volume. To set the upper bound, we reasoned that the 20% error in contour length could compound when we measured volume, so we set the bound at 1.7 to account for this while also not counting toroids that contain multiple DNA molecules. In addition, when we plotted the calculated volumes of 398-nm-length toroids, we observed that these bounds encompass a group of toroids clustered around a peak near the expected volume (Supplementary Figure S2).

### Calculating Average Loop Diameter

To calculate the average loop diameter for each DNA length, we first identified the peak in the toroid height histogram (Figure 6A; 1.5 nm for *L*=217 nm, 2.0 nm for *L*=398 nm, 4.0 nm for *L*=217 nm). Then, we divided this by the height of bare DNA on mica (0.5 nm) (1) to compute the number of loops in each of these toroids (3 for *L*=217 nm, 4 for *L*=398 nm, 8 for *L*=217 nm). After that, we made the assumption that all loops on the same molecule have the same circumference and that the entire length of the DNA is packaged into the toroid. We then computed the average circumference for a loop under these assumptions (72 nm for *L*=217 nm, 100 nm for *L*=398 nm, 128 nm for *L*=217 nm).

Next, we use a calibration curve to turn our loop circumference measurements into $\sigma_x$ values. Specifically, we use a calibration curve (Supplementary Figure S6) to calculate $\sigma_x$ at each number of loops and each DNA length. These $\sigma_x$ values are marked on our $\sigma_x$ peaks histograms (Figure 4B).

**Geometric Limits on Loop Diameter**

We wanted to determine whether loop sizes are constrained by geometry. To do this, we derived three models that set geometric limits on the loop diameter. Each of these models makes no reference to protamine-DNA interactions, instead making purely geometric arguments. For all models, we note that the lower limit on loop size is the pixel size in our AFM images (3.9 nm).

In the first model, we set an absolute upper bound on loop size given the DNA length. Here, we assume an *n*-loop flower and that no loops can be larger than the DNA contour length *L* minus (*n* - 1) loops that are all the minimum loop circumference (3.9 nm):

$$d_{max,absolute} = \frac{L-(3.9\,nm)\pi(n-1)}{\pi}. \qquad \text{(Equation S3)}$$

In the second model, we assume that all DNA is packaged within a loop and that all loops are the same size. We call this the uniform loops model. Under these assumptions, the loop diameter is

$$d_{max,uniform\,loops} = \frac{L}{n\pi}. \qquad \text{(Equation S4)}$$

In the third model, we also assume that all DNA is packaged within a loop, but now loops can be different sizes. Specifically, one loop can be larger than the others by the heterogeneity factor *H* (maximum measured loop diameter divided by minimum measured loop diameter in DNA molecules that only form a single loop). We call this model the nonuniform loops model. Under these assumptions, the maximum loop size is

$$d_{max,\,nonuniform\,loops} = \frac{L}{[1+(n-1)/H]\pi}. \qquad \text{(Equation S5)}$$

We observed in prior data collected on 105-nm DNA that the ratio of the diameters of the largest protamine-induced single loops to the smallest is about 1.75 (2), so we set $H$ = 1.75.

**Toroid Packing Models**

Hexagonal Packing

DNA is arranged in an ideal hexagonal lattice within the toroid. We assume that loops are added one-by-one. There is a discrete set of possible (diameter, height) pairs that a hexagonally packed toroid can have. We assume that the first loop has a diameter of 30 nm, which is in line with our observations of the average protamine-induced DNA loop (2). Each additional loop that expands the toroid horizontally has a diameter of the previous loop plus twice the thickness of DNA (4 nm). The diameter of the toroid *d* is then the diameter of the outermost loop. Each additional loop that is stacked vertically atop a previous loop adds 0.5 nm to the total height *h* of the toroid. We determined the complete set of possible (*d, h*) pairs for toroids with up to 10 loops by enumerating the possibilities (Supplementary Table S1).

For each length, we computed the maximum number of 30-nm-diameter loops and used this information to construct the set of possible points. These points bound a region that defines the hexagonal packing model.

Horizontal Packing Model

The horizontal packing model is a limiting case in which the toroid grows exclusively horizontally. In the horizontal packing model, loops are arranged concentrically. The innermost and outermost loops each have one lattice contact, while all other loops have two lattice contacts. Additional concentric loops are added to the outside of the toroid. In this model, the height is the height of a single DNA molecule, $a$, regardless of how many loops there are. The height for a horizontally packed toroid is therefore:

$$h = a. \qquad \text{(Equation S6)}$$

We use $a$=0.5 nm, in line with our observations of the height of bare DNA on mica (1).

Vertical Packing Model

The vertical packing model is a limiting case in which the toroid grows exclusively vertically. DNA loops stack directly on top of one another. All loops have the same diameter, so each additional loop increases the height of the toroid by $a$, but does not change its diameter. If the entire contour length $L$ of the DNA is used to form the toroid, then there are a total of $\frac{L}{\pi d}$ loops in the toroid. The height of the toroid is therefore:

$$h = \frac{La}{\pi}\left(\frac{1}{d}\right). \qquad \text{(Equation S7)}$$

We plot Equation S7 in Figure 6B, using $a$ = 0.5 nm and include a 20% error margin on this parameter. We set $L$ using the average experimentally measured value in non-toroid singlets in AFM data. We did not include an error margin on $L$. Observe that we can rewrite Equation S7 as:

$$\frac{h}{L} = \frac{a}{\pi}\left(\frac{1}{d}\right). \qquad \text{(Equation S8)}$$

We use this form in Figure 6C in order to pool datasets from multiple different DNA constructs and assess their agreement with the vertical packing model.

**FIGURES**

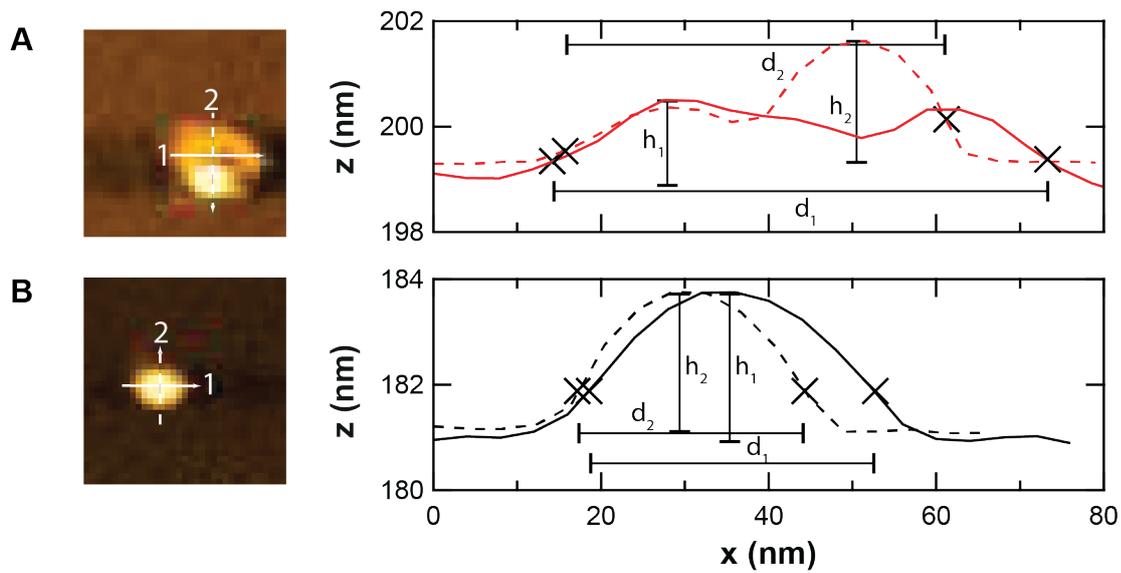

**Figure S1: Toroid diameter and height.** We take two perpendicular height scans across a toroid and then plot them. To identify the diameter, we calculate every local maximum of each graph and then locate the points at 1/e times this maximum (*black crosses*). The diameter is the maximum horizontal between any two of these points. This procedure is repeated for both profiles, and the diameter we report, *d*, is the average of the two measurements $d_1$ and $d_2$. To measure the toroid height *h*, we calculate the maximum toroid height relative to the baseline for both profiles and then average them. Finally, if a DNA molecule has two local maxima in at least one of the profiles, we say that this toroid has a hole, if not it is without a hole. Example toroids (A) with a hole and (B) without a hole are shown. All images are 395 nm x 395 nm. DNA is 398 nm long.

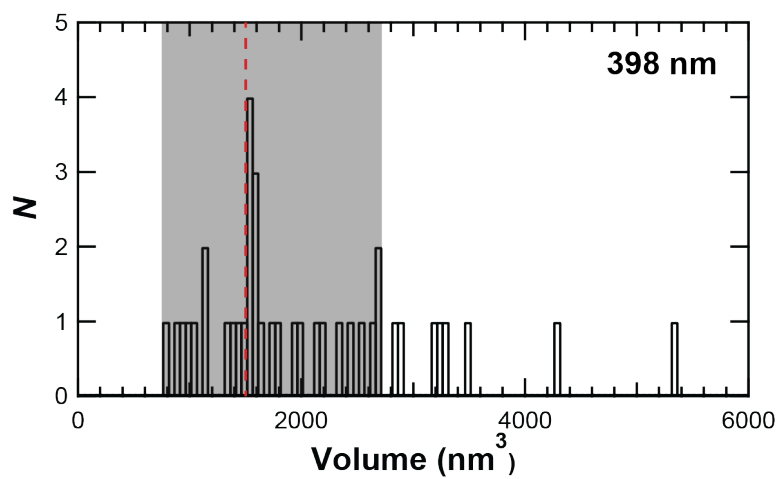

**Figure S2: Toroid volume cuts.** To identify valid toroids for analysis, we plotted the measured volume of all toroids for a particular DNA length (in the example above, $L$=398 nm). We then calculated the volume of the toroid assuming vertical stacking of loops in the toroid (*red dashed line*). We set an error tolerance of 0.5-1.7 times the expected volume (*grey box*). Toroids outside of this range were not analysed.

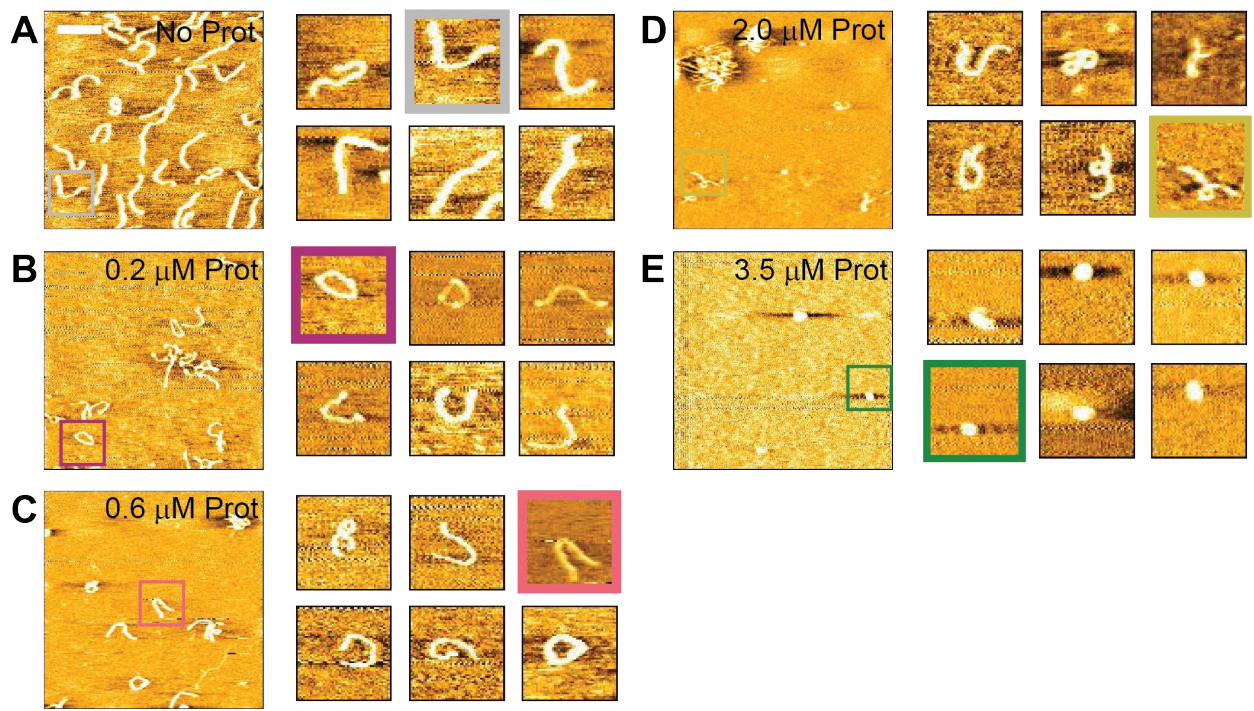

**Figure S3: Sample images for 217-nm-length DNA.** A-E) Representative 1 μm x 1 μm AFM height scans of 217-nm-length DNA (*left*) and 200 nm x 200 nm representative singlets (*right*) at each protamine concentration are shown. Coloured boxes indicate singlets extracted from the 1 μm x 1 μm AFM height scan shown. Scale bar is 200 nm.

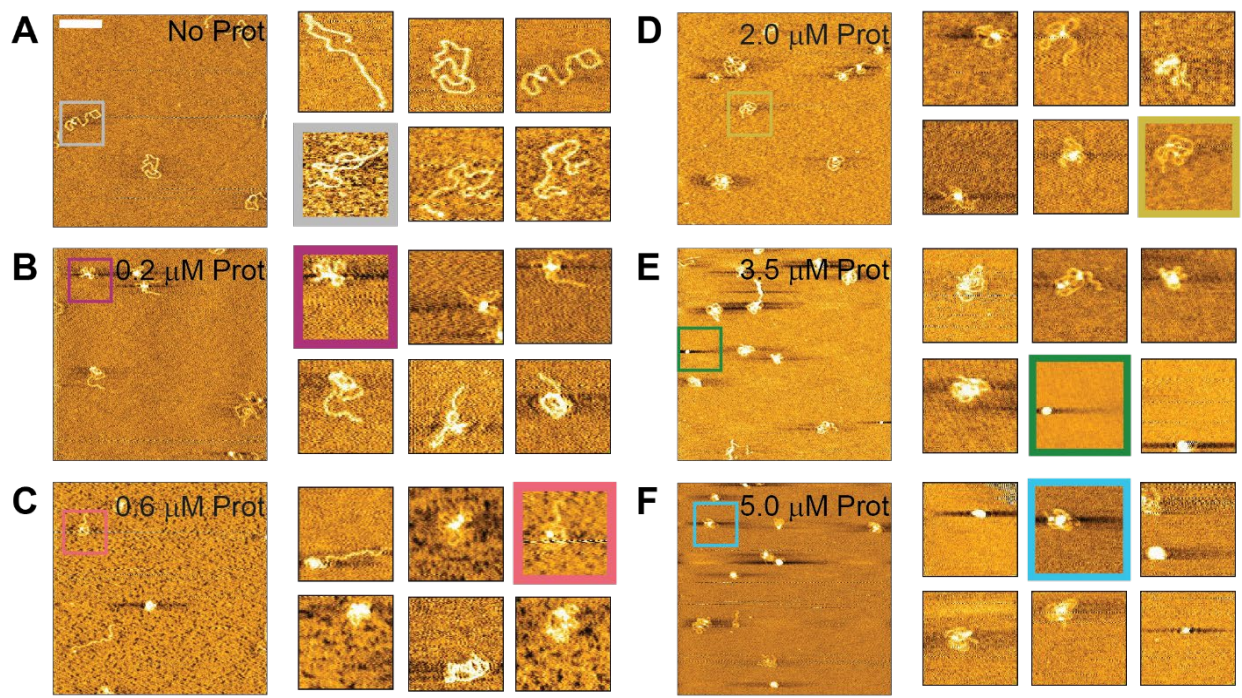

**Figure S4: Sample images for 1023-nm-length DNA.** A-F) Representative 2 μm x 2 μm AFM height scans of 1023-nm-length DNA (*left*) and representative singlets (*right*) for 1023-nm-length DNA at each protamine concentration are shown. Coloured boxes indicate singlets extracted from the 2 μm x 2 μm AFM height scan shown. Scale bar is 400 nm.

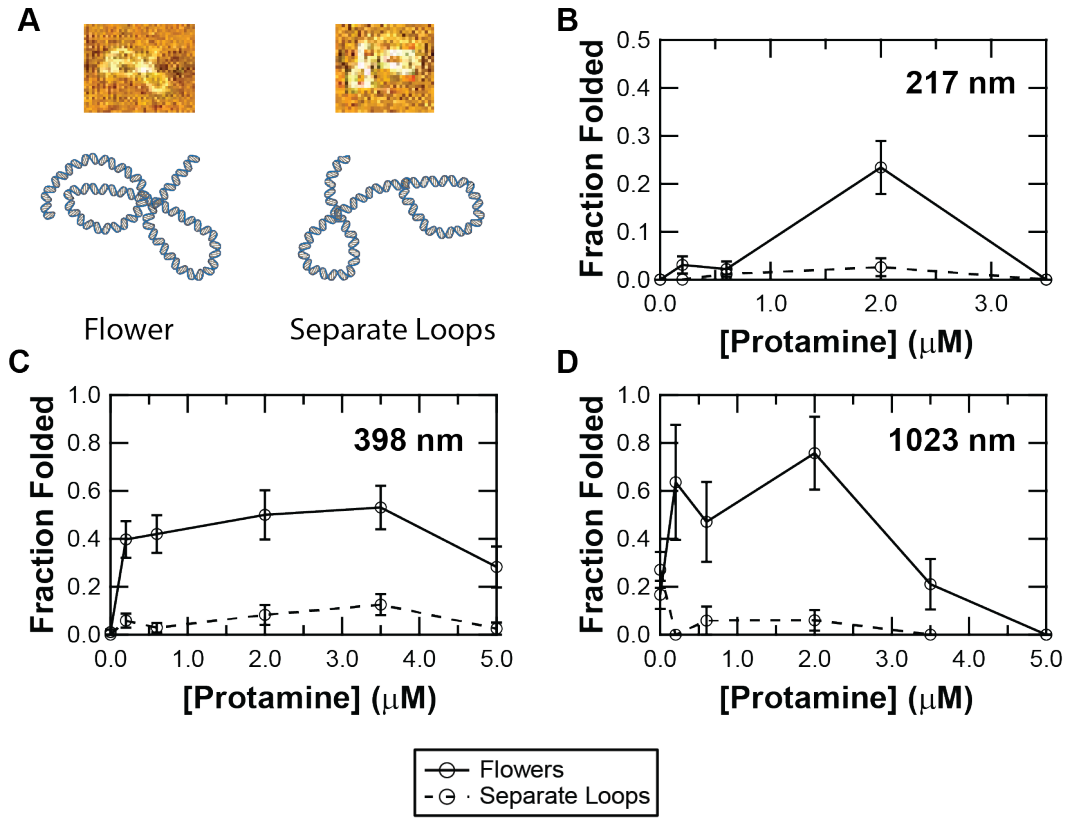

**Figure S5: Distinguishing between flowers and separate loops.** A) We distinguish between flowers, which have multiple loops that share a common centre, and separate loops, which have multiple, spatially separate loops. The fraction of singlets in either a flower or multiloop conformation at each protamine concentration for B) 217-nm-length, C) 398-nm-length, and D) 1023-nm-length DNA. The fraction of flowers is higher under all experimental conditions, in some cases by a factor of 10.

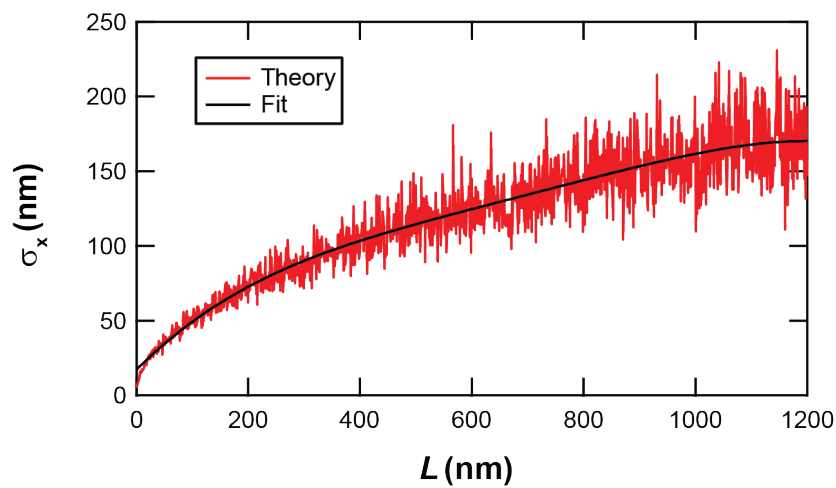

**Figure S6: TPM calibration curve.** We used a computer simulation to estimate the 50-frame rolling standard deviation $\sigma_x$ in a TPM experiment using a polymer with persistence length 40 nm (3) and a 560-nm-diameter particle in which the polymer's contour length $L$ ranged from 1-1200 nm, then fit a 5$^{th}$-degree polynomial to the simulation data. We use this polynomial to convert between $\sigma_x$ and $L$.

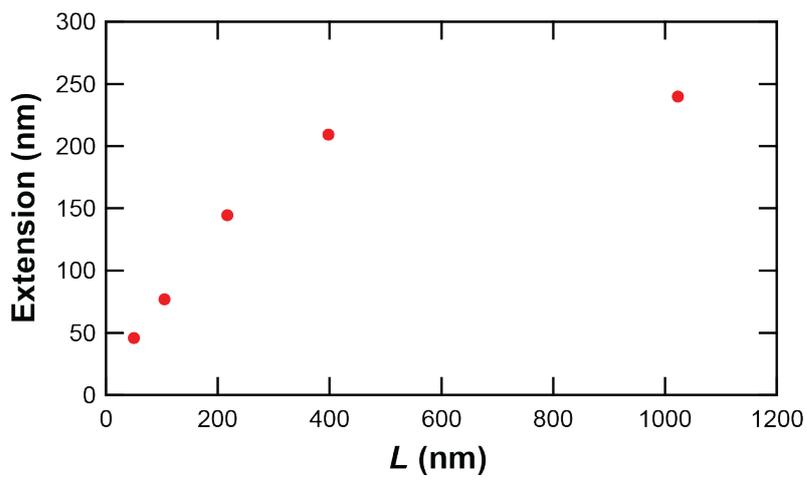

**Figure S7: Extension versus contour length for AFM singlets.** Average extension of unfolded molecules in the AFM assay at each DNA length, $L$. This data can be used as a calibration between extension and DNA length.

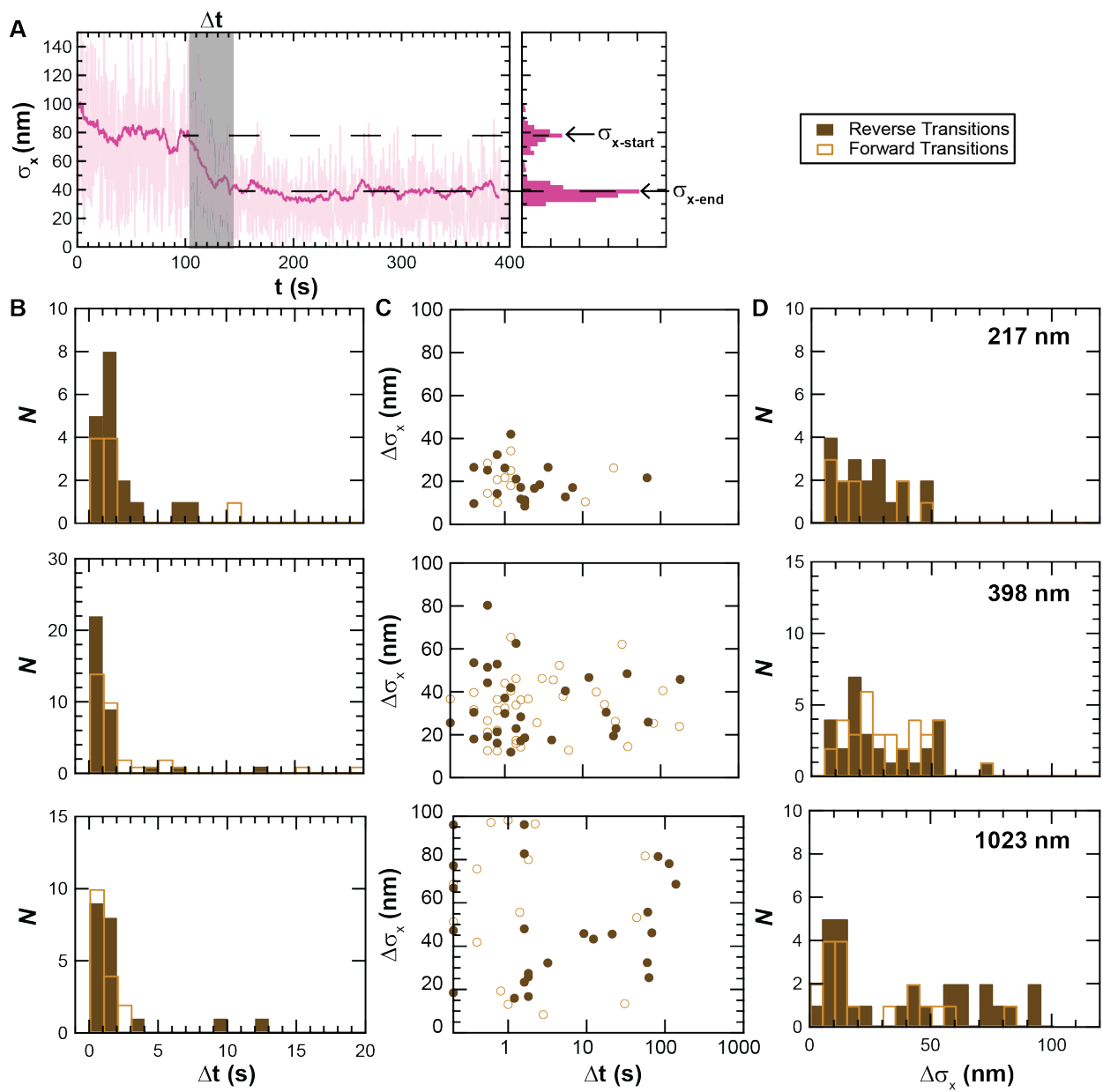

**Figure S8: Transition dynamics and reversibility.** A) Example of a transition for a 398-nm-length tether at 0.03 µM protamine. We determine the start and end standard deviations using a 50-frame rolling window (*magenta*) by identifying two distinct peaks in the histogram (*right*). We compute the duration of the transition Δ*t* using the standard deviation with a 2-point rolling window (*pink*). B) Histograms of durations for all forward (*hollow bars*) and reverse (*solid bars*) transitions in 217-nm-length (*N*=29, *top*), 398-nm-length (*N*=64, *middle*), and 1023-nm-length (*N*=45, *bottom*) DNA. The majority of transitions occur on a timescale of seconds or faster. 76% of transitions in 217-nm-length DNA, 66% of transitions in 398-nm-length DNA, and 67% of transitions in 1023-nm-length DNA are faster than 2 seconds. Bin size is 1 s. C) We plot the magnitude of the standard deviation change for each transition versus the duration of the transition (*reverse: solid brown circles, forward: hollow tan circles*) and find that the transition rates span three orders of magnitude for both 217-nm-length and 398-nm-length DNA. The x-axis is on a logarithmic scale. D) In comparing the end σ$_x$ of forward transitions with the start σ$_x$ of reverse transitions,

we find that both span the same range in all three DNA constructs examined. We did not include transitions that had $\sigma_x$ outside of our calibration range (Supplementary Figure S6).

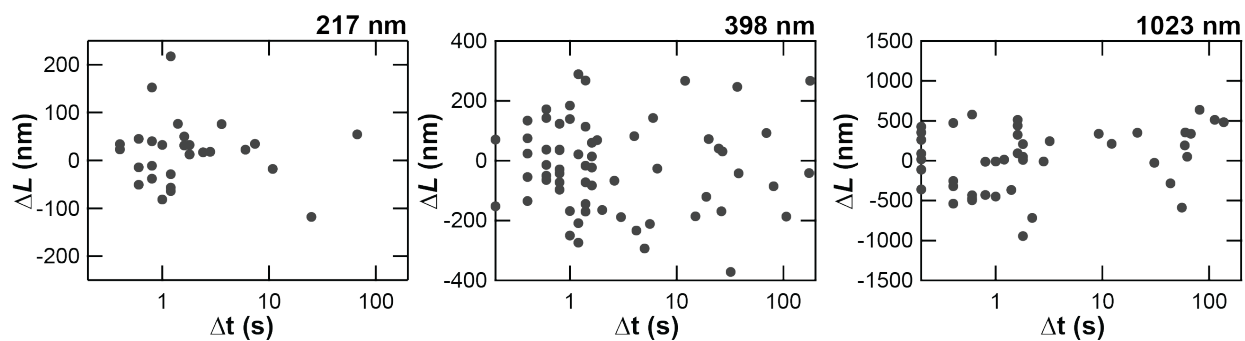

**Figure S9: Transition change in length versus time.** For each transition shown in Supplementary Figure S8, we computed the signed change in contour length $\Delta L$ using the calibration curve in Supplementary Figure S6. We then plotted this data set against the associated transition times $\Delta t$. Data are approximately symmetric about the $\Delta L = 0$ line, indicating that forward and reverse transitions have nearly identical dynamics. Note that the x-axis is on a logarithmic scale.

**TABLES**

| Number of Loops | d (nm) | h (nm) |
|---|---|---|
| 1 | 30 | 0.5 |
| 2 | 30 | 1 |
| 2 | 34 | 0.5 |
| 3 | 34 | 1 |
| 3 | 38 | 0.5 |
| 4 | 38 | 1 |
| 4 | 34 | 1.5 |
| 5 | 38 | 1.5 |
| 6 | 38 | 1.5 |
| 7 | 38 | 1.5 |
| 8 | 38 | 1.5 |
| 8 | 42 | 1.5 |
| 8 | 38 | 2 |
| 9 | 38 | 2 |
| 9 | 42 | 1.5 |
| 9 | 38 | 2.5 |
| 10 | 42 | 2 |
| 10 | 42 | 1.5 |
| 10 | 46 | 1.5 |
| 10 | 38 | 2.5 |

**Table S1: Height-Diameter Pairs for the hexagonal packing model**